\documentclass[preprint,aps,nofootinbib]{revtex4}
\usepackage{graphicx}
\usepackage{amsmath}
\usepackage{amsfonts}
\usepackage{amssymb}
\usepackage{color}%
\providecommand{\U}[1]{\protect\rule{.1in}{.1in}}

\newcommand{\f}{\begin{equation}}
\newcommand{\ff}{\end{equation}}
\newcommand{\fa}{\begin{eqnarray}}
\newcommand{\ffa}{\end{eqnarray}}

\begin{document}
\title{Bouncing universe from a modified dispersion relation}
\author{Yi Ling}
\email{yling@ncu.edu.cn}
\author{Wei-Jia Li}
\email{li831415@163.com}
\author{Jian-Pin Wu}
\email{jianpinwu@yahoo.com.cn} \affiliation{ Center for
Relativistic Astrophysics and High Energy Physics, Department of
Physics, Nanchang University, 330031, China}

\begin{abstract}

In this paper we argue that modified Friedmann equations with a
bounce solution can be derived from a modified dispersion relation
by employing a thermodynamical description of general relativity
on the apparent horizon.

\end{abstract}
\maketitle

\section{Introduction}
One remarkable achievement of general relativity is linking the
geometric structure of spacetime with the distribution of  matter
sources. Or in another word, the nature of matter sources determines
which kind of spacetime geometry we may obtain through the Einstein
equation. Once people know in first principle all the matter sources
should be described by quantum theory, it is realized that the
geometry of spacetime should be quantized as well. Before a complete
and consistent quantum theory of gravity can be established, one
widely accepted belief is that the combination of general relativity
and quantum mechanics will provide a fundamental minimal length or
maximal energy that can be detected or measured, which is always
termed as Planck length or Planck energy. If this is true, then it
is expected that the longstanding cosmological singularity problem
may be solved in the quantum theory of gravity. Semi-classical
theory of quantum gravity provides such an intuitive picture to
replace the cosmological singularity in standard cosmology by a big
bounce\cite{Bojowald1,Ashtekar1}, which is implemented by
considering modifications due to the quantum corrections to the
standard Friedmann equations at the semi-classical limit. Recently,
the big bounce as a solution to the modified Friedmann equations has
been extensively investigated in various
approaches\cite{Shtanov,Brown,Ashtekar2,Bojowald2,Casadio,Setare,Ding,Battisti,Wu1,
Kiritsis,Yifu,Mielczarek}. Moreover, the bounce solutions even exist
in some special models in the Einstein gravity
\cite{Starobinsky,Page,LW}. In this paper we intend to drive the
modified Friedmann equations with a big bounce solution from a
modified dispersion relation in a heuristic manner.

Our motivation directly comes from the fact that modified dispersion
relations may provide a bound for both the energy and the momentum
of particles, and such effects have greatly been studied on the fate
of Lorentz symmetry at extremely high energy level, as well as
quantum gravity
phenomenology\cite{Amelino03ex,Amelino03uc,Magueijo1}. Based on the
spirit of general relativity, if all the matter sources are
described by modified dispersion relations which prevent the energy
density from diverging at high energy level, then the curvature of
the corresponding spacetime should also be bounded such that it is
singularity free. Previously the quantum gravity effects of
particles on spacetime have been investigated in gravity's rainbow
formalism\cite{Magueijo2,Galan,Ling1,Hackett05mb,Wu1,Aloisio05qt,Galan06by,Ling06ba,Ling06az,Leiva}.
In this paper we intend to derive a modified Friedmann equation with
a bounce solution from a modified dispersion relation by employing
the thermodynamical description of the cosmological equations on the
apparent horizon of the universe.

Originally the thermodynamical description of the Einstein equation
is proposed by Jacobson\cite{Jacobson}. In this way the Einstein
equation is an equation of state and can be derived from a
fundamental thermodynamical relation, namely Clausius relation
$\delta Q=TdS$, which connects heat, entropy and temperature for all
the local Rindler causal horizons, where $\delta Q$ and $T$ are
interpreted as the {\it energy flux} and the Unruh temperature
detected by the accelerated observer inside the horizon,
respectively. Recently, this proposal has been testified in various
gravity theories and cosmological
models\cite{Padmanabhan,Akbar,Eling,Wang,Ge,Wu2,Gong}. In this
context, it turns out that the apparent horizon plays an appropriate
role in deriving the Friedmann equations from Clausius relation.
Subsequently, an analogy of the first thermodynamical law can also
be established on the apparent horizon. In particular, a corrected
entropy-area relation may give rise to a modified Friedmann equation
has been studied in \cite{Cai,Zhu}. However, along this direction
deriving a modified equation with a bounce solution has not been
succeeded in all previous papers. In this paper we argue that with
the help of modified dispersion relations
 this can be realized by applying it to the energy
flux passing through the apparent horizon.

Our paper is organized as follows. In next section we propose a
specific form of modified dispersion relation in which both energy
and momentum are bounded at the order of Planck scale. Then in
section three we derive the modified Friedmann equations from the
Clausius relation on the apparent horizon based on this modified
dispersion relation. The corresponding entropy-area relation is also
obtained with a logarithmic correction term. The conclusion and
discussion are given in section four.

\section{a modified dispersion relation }
In this paper we propose a modified dispersion relation at quantum
gravity phenomenological level, which has a form
\begin{equation}\label{B}
\frac{1}{\eta\l_{p}}\sin(\eta\l_{p}E)=\sqrt{p^{2}+m_{0}^{2}},
\end{equation}
where $E$ and $p$ are the energy and momentum of a particle with
mass $m_0$ respectively, and $\eta$ is a dimensionless parameter.
The Planck length $l_{p}$ as well as the Planck mass $M_{p}$ is
defined as $l_{p}\equiv\sqrt{8\pi G}\equiv1/M_{p}$ with $\hbar=c=1$.
Obviously, to keep the magnitude of the momentum positive definite,
we require that the energy $E$ takes the value in the range $[0 ,
\pi/ \eta\l_{p}]$. First of all, at low energy limit with $E\ll
M_p$, the Taylor expansion of this relation leads to
\begin{equation}\label{C}
(1-\eta^2\l_{p}^{2}E^{2}/3)E^{2}-p^{2}=m_{0}^{2},
\end{equation}
which is nothing but the standard dispersion relation when
$l_p\rightarrow 0$. The impacts of the deformed dispersion relation
with form Eq.(\ref{C}) on black hole thermodynamics have been
extensively studied in \cite{Ling1,Ling2,Han}. Now to obtain the
modified Friedmann equations with bounce solutions, we insist to use
the deformed dispersion relation Eq.(\ref{B}) rather than
Eq.(\ref{C}), and the reasons to do so will be presented later.

One  feature of the modified dispersion relation Eq.(\ref{B}) is the
peculiar relation between the energy and the momentum at Planck
energy level. From (\ref{B}) we easily see that in the energy range
of $[0 , \pi/ 2\eta\l_{p}]$, as usual the energy $E$ is a
monotonously increasing function of the momentum $p$. In particular
as $E=\pi/ 2\eta\l_{p}$, the momentum $p$ reaches its maximal value
$1/\eta\l_{p}$. However, as the energy climbs up further, the
momentum $p$ decreases monotonously until it reaches zero at $E=\pi/
\eta\l_{p}$. This property can become more transparent when we
consider the differential relation of the energy and momentum, which
can be derived as
\begin{equation}\label{D}
\delta E=\pm \frac{1}{\sqrt{1-\eta^2l_{p}^{2}p^{2}}}\delta p,
\end{equation}
where the positive sign $``+"$ corresponds to the energy range $[0
, \pi/ 2\eta\l_{p}]$, while the negative sign $``-"$ to $[\pi/
2\eta\l_{p} , \pi/ \eta\l_{p}]$. We point out that both cases are
essential to derive the modified Friedmann equations with bounce
solutions, especially to control the universe evolution around the
bouncing point, and this is what we intend to do in next section.

\section{modified Friedmann equation from thermodynamical description of GR}
Firstly we briefly review the thermodynamical description of the
ordinary Friedmann equations in cosmology. Without loss of
generality in this paper we only consider the spatially flat
universe. Given the flat Friedmann-Robertson-Walker metric
\begin{equation}\label{E}
  d{s}^{2}=-d{t}^{2}+{a}^{2}(t)(d{r}^{2}+r^2d\Omega^{2})
          =h_{ab}dx^{a}dx^{b}+\tilde{r}^{2}d{\Omega}^{2},
\end{equation}
where $x^{0}=t$ , $x^{1}=r$, $\tilde{r}=a(t)r$,
$h_{ab}=\textit{diag}(-1,a^2)$ and $d{\Omega}^{2}$ is the metric of
a 2-dimensional sphere with unit radius. Then it is straightforward
to obtain the radius of the apparent horizon which is defined as
$h^{ab}\partial_a\tilde{r}\partial_b\tilde{r}=0$. For the spatially
flat universe, it turns out that
\begin{equation}\label{F}
\tilde{{r}}_{A}=\frac{1}{H},
\end{equation}
which coincides with the Hubble horizon. Recently a lot of work
shows that an analogy of the first law of thermodynamics could be
constructed on the apparent horizon if one assumes that the apparent
horizon has an associated entropy $S$ and temperature $T$ which are
respectively identified as

\begin{equation}\label{G}
S=\frac{A}{4G},\ \ \ \ \ \ T=\frac{1}{2\pi \tilde{{r}}_{A}},
\end{equation}
where $A=4\pi \tilde{r}_{A}^2$ is the area of the apparent horizon.
More explicitly, the Friedmann equation on the apparent horizon can
be rewritten as the Clausius relation
\begin{equation}
\delta Q=TdS.
\end{equation}
Conversely, starting from the Clausius relation on the apparent
horizon, one can derive the standard Friedman equation
\begin{equation}\label{H} H^{2}=\frac{8\pi G}{3}\rho,
\end{equation}
through an integration\cite{Cai}
\begin{equation}\label{I}
 \frac{8\pi G}{3}\rho=-\frac{\pi}{G}\int
 S'(A)(\frac{4G}{A})^{2}dA,
\end{equation}
where $S'=\frac{dS}{dA}$ defines the variation of the entropy with
the area of the apparent horizon . In this case it is nothing but a
constant which is $\frac{1}{4G}$.

Next we consider how the modified dispersion relation will give rise
to a modified Friedmann equation by employing the heuristic analysis
which has been previously applied to Hawking radiation of black
holes. A similar consideration for the $FRW$ universe has also been
presented in \cite{Zhu}. Consider a single radiation particle with
quantum energy $\epsilon$ passes through the apparent horizon, then
from the Clausius relation we obtain the {\it minimal} change of the
entropy corresponding to the minimal change of the horizon area as
\begin{equation}\label{J}
  \delta S_{min}\sim dS=\frac{1}{T}\delta Q\simeq \frac{1}{T}\frac{1}{\delta x},
\end{equation}
where we have employed the standard dispersion relation and
Heisenberg's uncertainty relation, namely identifying the changes
$\delta Q\sim \epsilon\sim \delta E\sim \delta p\sim \frac{1}{\delta
x}$. Furthermore, we identify the uncertainty of particle position
as its Compton wavelength which is inversely proportional to its
Hawking temperature, namely $\delta x\sim \tilde{r}_{A}$
\cite{Adler,Bolen,Nozari}. Given the assumption that the minimal
change of the horizon area $\delta A$ of a quantum system is
$l_p^2=8\pi G$\cite{Bekenstein,Maggiore:2007nq,Medved:2009nj}, we
find that
 \f S'={dS\over
dA}\cong {\delta S_{min}\over \delta A_{min}}={1\over
4G},\label{S1}\ff leading to the standard Bekenstein-Hawking
entropy. It is worthwhile to point out that the analysis based on
the thermodynamical description here obtains the same result as the
one based on the Bekenstein entropy assumption for black holes, as
discussed in \cite{Ling2}.

Now if we take the modified dispersion relation into account,
using Eq.(\ref{D}) we find the minimal change of the entropy is
corrected as
\begin{equation}\label{K}
  \delta S_M^{min}\sim dS_{M}\simeq \pm\frac{1}{\sqrt{1-\frac{4\pi \eta^2
   l_{p}^{2}}{A}}}\frac{1}{T}\frac{1}{\delta
  x}\equiv \pm 2\pi f_{M}(A),
\end{equation}
where we have set $p\sim \delta p\sim {1\over \delta x}\sim
{1\over \tilde{r}_A}$ since the maximal momentum uncertainty of a
quantum particle $\delta p$ is of order of $p$, which will lead to
the minimal change of the entropy \cite{Adler,Ling2,Han}. Thus
 \f S'={dS_M\over
dA}\cong {\delta S_{min}\over \delta A_{min}}=\pm{1\over
4G}f_{M}(A).\label{S2}\ff

Since the positive sign $``+"$ corresponds to the energy range $[0
, \pi/ 2\eta\l_{p}]$, which covers all the classical and
semi-classical region below the order of Planck energy, we
consider it firstly. Integrating Eq.(\ref{S2}) we obtain the
entropy-area relation with corrections due to the modified
dispersion relation as
\begin{equation}\label{L}
S_{M}=\frac{A}{4G}\sqrt{1-\frac{4\pi\eta^2 l_{p}^{2}}{A}}+
\frac{\pi\eta^2 l_{p}^{2}}{2G} \ln[A+A\sqrt{1-\frac{4\pi\eta^2
l_{p}^{2}}{A}}-2\pi\eta^2 l_{p}^2]+C,
\end{equation}
where an integral constant $C$ has been set to $4\pi^2 \eta^2 (1-\ln
2)$ simply requiring that $S_{M}$ has a classical limit
$\frac{A}{4G}$ as $A\gg 4\pi \eta^2 l_{p}^{2}$. Notice that in this
formalism the correction term has a logarithmic form but the factor
is $4\pi^2\eta^2$ which is always positive, different from the
previous results for entropy correction of black
holes\cite{Ling2,Medved,Camelia,Myung}. On the other hand, when the
area $A$ is small enough to be comparable with the Planck scale, the
famous factor $1/4$ in Bekenstein-Hawking entropy formula is
changed. Moreover, we find that the area of the apparent horizon is
bounded, namely $\tilde{r}_A\geq \eta l_p$, implying a finite
temperature on the horizon even at the Planck scale.

Now given a corrected entropy-area relation, one can obtain the
modified Friedmann equation by substituting the result of
Eq.(\ref{S2}) into Eq.(\ref{I})
\begin{equation}\label{O}
\frac{8\pi G}{3}\rho=-\frac{2}{\eta ^2
l_{p}^{2}}\sqrt{1-\eta^2l_{p}^{2}{H}^{2}}+C_{1},
\end{equation}
where $C_{1}$ is an integral constant we need to set. Obviously,
we require that this equation should return to the standard
Friedmann equation  at low energy limit $l_{p}\rightarrow 0$. Thus
we have
\begin{equation}
C_{1}=\frac{2}{\eta^2l_{p}^{2}}-{\Lambda\over 3},
\end{equation}
where $\Lambda/3$ is a cosmological constant term independent of
$l_p$.
 Eq.(\ref{O}) can
be rewritten as
\begin{equation}\label{Q}
{H}^{2}=\frac{8\pi G}{3}\rho_t(1-\frac{\rho_t}{\rho_{c}}),
\end{equation}
with $\rho_{c}=12/\eta^2{l_{p}}^{4}$ and the total energy density
$\rho_t=\rho+\frac{\Lambda}{8\pi G}$. Subsequently using the
conservation equation $\dot{\rho}+3H(\rho+P)=0$, one has
\begin{equation}\label{R}
\dot{H}=-4\pi G(\rho+P)(1-2\frac{\rho_t}{\rho_{c}}).
\end{equation}
 The above two equations can also be
obtained in semiclassical loop quantum cosmology with holonomy
corrections and some effective
theories\cite{Ashtekar2,Bojowald2,Shtanov,Battisti}, where the
critical energy density $\rho_c$ has the same order as Planck
energy density. Indeed the big bounce solution to Eq.(\ref{Q})
exists whenever $H=0$ and $\dot{H}>0$, which implies
$\rho_t=\rho_c$. However, this solution is {\it not} contained in
Eq.(\ref{O}). This can be easily seen if one sets $H=0$ in
Eq.(\ref{O}), the unique solution to the energy density is
$\rho_t=0$.  As a matter of fact, in this equation $\rho_t$ only
values in the range of $[0 ,6/\eta^2{l_{p}}^{4}]$. So, in order to
obtain the bounce solutions, we need consider the differential
relation of energy and momentum at higher energy level which is
Eq.(\ref{D}) with a
 $``-"$ sign. Repeat the derivation above, we obtain
the modified Friedman equation as
\begin{equation}\label{U}
\frac{8\pi
G}{3}\rho=\frac{2}{\eta^2l_{p}^{2}}\sqrt{1-\eta^2l_{p}^{2}{H}^{2}}+C_{2},
\end{equation}
where the constant $C_{2}$ can be determined by keeping $H^2$ as a
single-valued and continuous function of the energy density
$\rho_t$. Namely, at $H^2=1/\eta^2l_p^2$, we should obtain the same
energy density through Eq.(\ref{O}) and Eq.(\ref{U}), which leads to
\begin{equation}\label{V}
 C_{2}=C_{1}=\frac{2}{\eta^2 l_{p}^{2}}-{\Lambda\over 3}.
\end{equation}
In Eq.(\ref{U}) $\rho_t$ only values in the range of
$[6/\eta^2{l_{p}}^{4},12/\eta^2{l_{p}}^{4}]$. But combining
Eq.(\ref{O}) and Eq.(\ref{U}) together, we find they cover all the
evolution range controlled by Eq.(\ref{Q}). Therefore, we conclude
that the modified Friedmann equations with bounce solutions can be
obtained from the modified dispersion relation presented in
Eq.(\ref{B}) through a thermodynamical description of general
relativity on the apparent horizon.

However, at the end of this section we need point out that the
entropy-area relation obtained in the energy density range of
$[6/\eta^2{l_{p}}^{4},12/\eta^2{l_{p}}^{4}]$ is problematic. As
from Eq.(\ref{S2}) we find the entropy would be
\begin{equation}\label{W}
S_{M}=-\frac{A}{4G}\sqrt{1-\frac{4\pi\eta^2
l_{p}^{2}}{A}}-\frac{\pi \eta^2 l_{p}^{2}}{2G}
\ln[A+A\sqrt{1-\frac{4\pi\eta^2 l_{p}^{2}}{A}}-2\pi\eta^2
l_{p}^2]+C_3,
\end{equation}
where constant $C_3$ is set to have the same entropy at
$\rho_t=6/\eta^2{l_{p}}^{4}$ through Eq.(\ref{L}) and Eq.(\ref{W}).
It turns out that
\begin{equation}\label{W}
C_3=4\pi^2\eta^2[1+2\ln (8\sqrt{2}\pi^2\eta^2)].
\end{equation}
Obviously, the entropy would be negative as the scale factor
approaches to the bounce point, where $H\rightarrow 0$ and $A
\rightarrow \infty$. This results from the peculiar differential
relation between the energy and the momentum at Planck scale in the
modified dispersion relation, implying that when evolving back to
the big bounce, the entropy continuously decreases while the area of
the apparent horizon will increase after reaching its minimum value
$\tilde{r}_A=\eta l_p$. This sounds very unreasonable, if one
insists that entropy is defined as a measure of the number of
microscopic quantum states. Previously negative entropy also
appeared in the context of black holes and more recently phantom
cosmology, and some relevant discussions can be found for instance
in\cite{Pollock:1989pn,Cvetic:2001bk,Brevik:2004sd,Izquierdo:2005ku,Izquierdo:2006tp}.
We think the appearance of negative entropy near the bounce point is
analogous to the situation in phantom dominated universe, since the
null energy condition is violated in the classical framework with a
bounce universe. In our paper the presence of negative entropy
strongly implies that one need find a new way to define the entropy
of the universe near the bounce point. As a matter of fact, the
entropy problem is a well-known open question in bouncing or cyclic
universe scenarios if we respect the second law of the
thermodynamics, stating that the total entropy of the universe
should not decrease forever. Then the entropy should not have a
proportional relation with the area of the apparent horizon for a
contracting universe or any phase during which the area of the
apparent horizon is decreasing, for instance after the bounce point
of the universe. Our picture implies that the Clausius relation
maybe has not a thermodynamical interpretation when the system is
far from the equilibrium state, or the entropy in such a
thermodynamical relation has no longer a statistical interpretation.
In statistics, the appearance of negative entropy usually implies a
metastable state and some kind of phase transition could
occur\cite{Cvetic:2001bk}. We expect that near the big bounce point,
the universe should be described by a genuine quantum theory of
gravity, thus as the energy scale falling down, it might undergo a
phase transition from a quantum phase to semi-classical phase
through some decoherence mechanism.

\section{Summary and discussions}
In this paper we have proposed a specific modified dispersion
relation in which both energy and momentum of matter sources are
bounded, and with the assumption that the Clausius relation holds on
the apparent horizon, we have derived a modified Friedmann equation
with a big bounce solution. Comparing with previous work on the
modification of Friedmann equations from corrected entropy-area
relation or generalized uncertainty principle, the novelty in our
scenario can be summarized as the following two points. First of
all, those modified Friedmann equations do not contain a big bounce
solution. Mathematically we think the reason may be that the
correction terms taken into account in literature are just the
approximate expansion of all the corrections at some inappropriate
limit. An obvious example can also be seen in loop quantum
cosmology\cite{Bojowald2}. If one takes the limit of  the lattice
parameter $\mu$ vanishing and consider some low orders of the
expansion, as naively supposed in usual semi-classical approach,
then it is easily found that the resulted Friedmann equation does
not contain any bounce solution. The similar case happens in our
paper. If we started from the modified dispersion relation
Eq.(\ref{C}), then the big bounce solution could not be obtained.
Secondly, at Planck scale the scenario described by the modified
dispersion relation Eq.(\ref{B}) might be analogous to T-duality in
string theory\cite{BBS}. In the context of T-duality, it is believed
that there is a critical radius of the compactification, and physics
on length scales below this radius can equally well be described by
physics on length scales larger than that.  So the temperature is
also invariant under a T-duality transformation (see Brane Gas
Cosmology for example\cite{Alexander}). Similarly, like the critical
radius of the compactification in string theory, here the critical
size of the scale factor is reached at $\tilde{r}_A=\eta l_p$.
Before that moment, the temperature is increasing from $0$ to a
critical value
 $\frac{1}{2\pi \eta l_{p}}$. Inversely, in low energy region it
will decrease monotonously to be vanished. Moreover, if we obtain
the entropy of the universe after the bounce point, it is found that
the entropy is monotonously increasing such that the second law of
thermodynamics is respected, this can be seen from Eq(13). However
the magnitude of the entropy is negative near the bounce point, the
possible reason is that the semi-classical theory breaks down at so
high energy scale, and the universe should be described by a genuine
quantum gravity theory which is expected to be founded in future.

\section*{Acknowledgement}
Y. Ling is grateful to Prof. Pisin Chen for helpful discussion. This
work is partly supported by NSFC(Nos.10663001,10875057), JiangXi
SF(Nos. 0612036, 0612038), Fok Ying Tung Eduaction Foundation(No.
111008) and the key project of Chinese Ministry of
Education(No.208072). We also acknowledge the support by the Program
for Innovative Research Team of Nanchang University.

\end{document}